\documentclass[runningheads]{llncs}
\usepackage{url}

\usepackage{balance} % for \balance command ON LAST PAGE (only there!)
\usepackage[utf8]{inputenc}
\usepackage[T1]{fontenc}
\usepackage{multirow}
\usepackage{graphicx}
\usepackage{caption}
\usepackage{subcaption}

\newcommand{\reffig}[1]{Fig.~\ref{#1}}
\newcommand{\reftab}[1]{Table~\ref{#1}}
\newcommand{\refsec}[1]{Section~\ref{#1}}

\usepackage{subcaption}
\begin{document}
\title{A Comparative Analysis of Knowledge Graph Query Performance}

\titlerunning{A Comparative Analysis of Knowledge Graph Query Performance}
% If the paper title is too long for the running head, you can set
% an abbreviated paper title here
%
\author{Masoud Salehpour \and Joseph G. Davis}
\authorrunning{Masoud Salehpour and Joseph G. Davis}
% First names are abbreviated in the running head.
% If there are more than two authors, 'et al.' is used.
%
\institute{University of Sydney}
\maketitle % typeset the header of the contribution
\begin{abstract}
As Knowledge Graphs (KGs) continue to gain widespread momentum for use in different domains, storing the relevant KG content and efficiently executing queries over them are becoming increasingly important. A range of Data Management Systems (DMSs) have been employed to process KGs. This paper aims to provide an in-depth analysis of query performance across diverse DMSs and KG query types. Our aim is to provide a fine-grained, comparative analysis of four major DMS types, namely, row-, column-, graph-, and document-stores, against major query types, namely, subject-subject, subject-object, tree-like, and optional joins. In particular, we analyzed the performance of row-store Virtuoso, column-store Virtuoso, Blazegraph (i.e., graph-store), and MongoDB (i.e., document-store) using five well-known benchmarks, namely, BSBM, WatDiv, FishMark, BowlognaBench, and BioBench-Allie. Our results show that no single DMS displays superior query performance across the four query types. In particular, row- and column-store Virtuoso are a factor of 3-8 faster for tree-like joins, Blazegraph performs around one order of magnitude faster for subject-object joins, and MongoDB performs over one order of magnitude faster for high-selective queries.

\end{abstract}
%
%
%
%--------------------------------------------------
% Introduction
%--------------------------------------------------
\section{Introduction}
\label{sec::introduction}
The term Knowledge Graph (KG) was used by Google in 2012, referring to collecting information about real-world entities and their \textit{inter-relationships} to facilitate the exploitation of semantics for searching the Web. From a broader perspective, any labeled directed graph-based representation of knowledge in a particular domain can be called a KG~\cite{kgsurvey}. For example, the term KG has been used to refer to Semantic Web Linked Datasets such as DBpedia or YAGO. In the recent years, many organizations such as Amazon, Facebook, Microsoft, and Alibaba have created large KGs for different purposes ranging from \textit{semantic search}, \textit{recommendations}, \textit{reasoning}, and \textit{data integration}. However, unlocking KGs' full potential in response to the growing deployment requires data frameworks to represent KG content and data platforms that can efficiently store the content and execute queries over them.

For the data framework, the World Wide Web Consortium (W3C) has recommended the Resource Description Framework (RDF) as a directed and labeled graph-like structure for \textit{representation}, \textit{integration}, and \textit{exchange} of the content of a KG using a large set of triples of the form $<$subject predicate object$>$. RDF offers a simple representation in which subjects and objects of triples are vertices of a graph that are connected by predicates as labeled edges. This simplicity can help provide an intuitive conceptualization of real-world entities and their inter-relationships. It can also represent diverse KG content ranging from \textit{structured} to \textit{unstructured}. However, this \textit{flexibility} as well as the absence of an \textit{explicit schema} and the \textit{heterogeneity} of KG content pose a challenge to Data Management Systems (DMSs) for \textit{querying} KGs efficiently since DMSs typically cannot make any priori assumptions about the structure of the KG content~\cite{IBMapple}.

For the data platforms, DMS designers have employed a variety of \textit{design choices and architectures} to tackle the above-mentioned challenges for querying KGs. For example, a variety of exhaustive indexing strategies, compression techniques, and dictionary encoding (i.e., to keep space requirements reasonable for excessive indexing) have been implemented by major native RDF-stores such as multiple bitmap indexes of Virtuoso or dictionary-based lexical values encoding of Blazegraph. A number of research prototypes have also been presented. For instance,~\cite{tamer2019} proposed a workload-adaptive and self-tuning RDF-store using physical clustering of the underlying data and~\cite{RDF3x} followed a RISC-style (reduced instruction set) architecture to leverage multiple query processing algorithms and optimization. However, the problem of storing and querying KGs efficiently continues to challenge DMS designers.

In addition to the design choices and architectures of DMSs, KG query performance is also affected by the diversity of SPARQL \textit{query types}~\cite{watdivshort}. While the importance of these factors has been recognized, our understanding of the \textit{comparative} performance of different types of queries across the major DMS types is somewhat limited. In this paper, we explore this problem, including the \textit{interactions} between a DMS and \textit{query types}. We provide a fine-grained, comparative analysis of four major DMS types, namely, row-, \mbox{column-,} graph-, and document-stores, against major types of KG queries, namely, subject-subject (aka, star-shape), subject-object (aka, chain-like or path), tree-like (aka, combined), and optional (aka, left-outer-join or OPT clauses) join queries. The performance of row-, column-, and graph-stores for executing queries has been studied in~\cite{watdivshort} based on their widespread use for processing RDF data. A widely accepted typology of KG queries is yet to emerge. At this stage query types such as subject-subject, subject-object, tree-like, and optional queries have been analyzed in previous research. Query types such as subject-subject, subject-object, and tree-like have been the focus of experiments in~\cite{Survey2018short}.~\cite{Medha2} has highlighted the importance of optional queries.

For our experiment, we selected row-store Virtuoso, column-store Virtuoso, Blazegraph, and MongoDB as representative DMSs for row-, \mbox{column-,} graph-, and document-store, respectively. We loaded five well-known benchmark datasets, namely, BSBM, WatDiv, FishMark, BioBench-Allie, and BowlognaBench into the DMSs separately. The benchmark queries were executed over each of the DMSs separately and query execution times computed to analyze the effects of query types on the performance of different DMS types. Our contributions include:

\begin{itemize}
  
    \item Comparative performance analysis and experimental evaluation of row-, \mbox{column-,} graph-, and document-stores in supporting the different SPARQL query types %(including exploring the \textit{interactions} between \textit{DMS} and \textit{query types}
    
    \item Providing explanations for the observed strengths and limitations of the different DMSs depending on the types of queries %such as data locality, lower memory usage (i.e., because of lesser memory allocation to intermediate results). We also speculate that such locality leads to more optimal CPU cache (i.e., L2 cache) utilization.
    
    \item Communicating clear scientific and practical guidelines to researchers and practitioners through summarizing the lessons learned from our journey% and discussing some of the limitations.

\end{itemize}

The remainder of this paper is organized as follows. In~\refsec{sec::preliminaries}, we provide some preliminary information about KG query types.~\refsec{sec::Experimental-Setting} presents our experimental setup including the KG benchmark characteristics, computational environment, DMSs configuration, indexing, and data loading process. In~\refsec{sec::Evaluation}, results of the query processing and related analyses are presented. We summarize the lessons learned from our research and discuss some of the limitations in \refsec{sec::Discussion}.~\refsec{sec::related-work} highlights related work. We present our conclusions and future work in~\refsec{sec::conclusion}.

\begin{figure}[t]
\centering\includegraphics[width=0.6\textwidth]{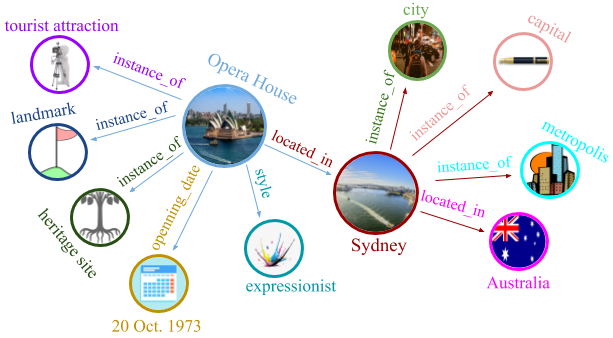}
\caption{An example of a simple Knowledge Graph describing the \mbox{``OperaHouse''}, a heritage site located in Sydney.}% Associated predicates and objects to each subject are in the same color.}%, e.g., the ``OperaHouse'' is a subject, ``located\_in'' is a predicate, and ``Sydney'' is an object.}
\label{fig::kg}
\end{figure}

\section{Query Types}
\label{sec::preliminaries}
In this section, we present some preliminary information about different query types using the \mbox{``OperaHouse''} KG example depicted in \reffig{fig::kg}. The content of this KG can be represented by the following triples\footnote{We use human-readable names in our examples in this paper.}:

\begin{verbatim}
  OperaHouse located_in    "Sydney"
  OperaHouse instance_of   "landmark"
  OperaHouse instance_of   "heritage site"
  OperaHouse instance_of   "tourist attraction"
  OperaHouse style         "expressionist"
  OperaHouse opening_date  "20 Oct. 1973"
  Sydney     located_in    "Australia"
  Sydney     instance_of   "city"
  Sydney     instance_of   "capital"
  Sydney     instance_of   "metropolis"
\end{verbatim}
%\bigskip

%This KG (see~\reffig{fig::kg}) contains 10 triples. For instance, \mbox{$<$OperaHouse located\_in "Sydney"$>$} is a triple in which \mbox{``OperaHouse''} is a subject, \mbox{``located\_in''} is a predicate, and \mbox{``Sydney''} is an object. 

An example of a query is then given below. It asks for the subject ``OperaHouse's'' architectural style name from the KG in \reffig{fig::kg}. ``?styleName'' is a variable to return the associated value (i.e., ``expressionist'') as the result. Queries may contain a set of triple patterns such as \textit{\mbox{``OperaHouse style ?styleName''}} in which the subject, predicate, and/or object can be a variable.

\begin{verbatim}
    SELECT ?styleName
    WHERE {
    OperaHouse style ?styleName .
    }
\end{verbatim}

Each triple pattern typically returns a subgraph. This resultant subgraph can be further \textit{joined} with the results of other triple patterns to return the final resultset. In practice, there are four major types of join queries: (i) Subject-subject joins (aka, star-like), (ii) subject-object joins (aka, chain-like or a path), (iii) tree-like (i.e., combination of subject-subject and subject-object joins), and (iv) optional joins (aka, left-outer-join or OPT clauses).

\textbf{Subject-subject joins.} A subject-subject join is performed by a DMS when a KG query has at least two triple patterns such that the predicate and object of each triple pattern is a given value (or a variable), but the subjects of both triple patterns are replaced by the \textit{same} variable. For example, the following query looks for all subjects of the KG in~\reffig{fig::kg} that are located in ``Sydney'' and their style is ``expressionist'' (the result will be ``OperaHouse'').

\begin{verbatim}
    SELECT ?x
    WHERE {
    ?x  style       "expressionist" .
    ?x  located_in  "Sydney"        .
    }
\end{verbatim}

\textbf{Subject-object joins.} A subject-object join is performed by a DMS when a KG query has at least two triple patterns such that the subject of one of the triple patterns and the object of the other triple pattern are replaced by the same variable. For example, the following query looks for all subjects that are located within Australian cities (the result will be ``OperaHouse'').

\begin{verbatim}
    SELECT ?y
    WHERE {
    ?x  located_in  "Australia"  .
    ?y  located_in  ?x           .
    }
\end{verbatim}

\textbf{Tree-like joins.} A tree-like joins consists of a \textit{combination} of subject-subject and subject-object joins. For example, the following query requires a tree-like join to look for the opening date of ``OperaHouse'' (the result will be ``20 Oct. 1973''). %In this example, two subject-subject joins and one subject-object join are combined. 

\begin{verbatim}
    SELECT ?y
    WHERE {
    ?x  opening_date     ?y        .
    ?x  located_in       ?z        .
    ?z  instance_of   "capital"    .
    ?z  instance_of  "metropolis"  .
    }
\end{verbatim}

\textbf{Optional Joins.} Queries return resultsets only when the entire query pattern matches the content of the KG. However, optional joins allow KG queries to return a resultset even if the optional part of the query is not matched since completeness and adherence of KGs' content to their formal ontology specification is not always enforced. This makes optional join a suitable tool for querying KGs. For example, the following query using optional join (in addition to a subject-subject join) to return ``OperaHouse'' as one of Sydney's tourist attractions.

\begin{verbatim}
    SELECT ?x
    WHERE {
    ?x  instance_of  "tourist attraction" .
    ?x  located_in   "Sydney"             .
    OPTIONAL {?x  instance_of  "zoo"      .}
    }
\end{verbatim}

\textbf{Selectivity of Queries.} As above mentioned, each KG query contains a set of triple patterns where a triple pattern is a structure of three components which maybe concrete (i.e. bound) or variable (i.e. unbound). Sets of triple patterns specify the complexity of the access to the underlying data. When the number of stored triples satisfying sets of triple patterns' conditions is large (i.e., as compared to the total number of stored triples), the corresponding query considered as low-selective~\cite{selectivity}. In this regard, each query type can also be either high-selective or low-selective depending on the number of stored triples satisfying its triple patterns' conditions.

\begin{table*}[t]
\centering
\begin{tabular}{c|l|r|c|r|rl}
\hline
 Benchmark & Scale (nominal) & \#Subjects  & \#Predicates & \#Objects  & \#Triples \\
 \hline\hline

\multicolumn{1}{ c|  }{\multirow{3}{*}{BSBM} } &
\multicolumn{1}{ c| }{10M} & 934,324 & 40 & 1,919,901 & 10,190,687    \\ %\cline{2-6}
\multicolumn{1}{ c|  }{}                        &
\multicolumn{1}{ c| }{100M} & 9,197,305 & 40 & 15,207,734 & 100,652,457    \\ %\cline{2-6}
\multicolumn{1}{ c|  }{}                        &
\multicolumn{1}{ c| }{1000M} & 91,647,129 & 40 & 140,996,171 & 1,004,406,629     \\ \cline{1-6}
\hline
\hline
\multicolumn{1}{ c|  }{\multirow{3}{*}{WatDiv} } &
\multicolumn{1}{ c| }{10M} & 521,585 & 86 & 1,005,832 & 10,916,457 \\
%\cline{2-6}
\multicolumn{1}{ c|  }{}                        &
\multicolumn{1}{ c| }{100M} & 5,212,385 & 86 & 9,753,266 & 108,997,714 \\ %\cline{2-6}
\multicolumn{1}{ c| }{}                        &
\multicolumn{1}{ c| }{1000M} & 52,120,385 & 86 & 92,220,397 & 1,092,155,948 \\ \cline{1-6}
\hline
\hline
\multicolumn{1}{ c|  }{\multirow{1}{*}{BioBench-Allie} } &
\multicolumn{1}{ c| }{100M} & 19,227,252 & 26 & 20,287,231 & 94,420,988    \\ \cline{1-6}
\hline
\hline
\multicolumn{1}{ c|  }{\multirow{1}{*}{FishMark} } &
\multicolumn{1}{ c|}{10M} & 395,491 & 878 & 1,148,159 & 10,002,178    \\ \cline{1-6}

\end{tabular}
\caption{Statistics of the Benchmark datasets}
\label{table::kgs}
%\vspace{-4mm}
\end{table*}

%****************************************
%       Experimental Setting
%****************************************
\section{Experimental Setup}
\label{sec::Experimental-Setting}
\noindent In this section, we present our experimental setup including the KG benchmark characteristics, computational environment, DMSs' configuration, indexing, and data loading process.

%########################################
%       Benchmark
%########################################
\subsection{Benchmarks}
We used five well-known benchmarks in this research. These are publicly available datasets along with a collection of queries. These benchmarks are: mBerlin SPARQL Benchmark\footnote{\url{http://wifo5-03.informatik.uni-mannheim.de/bizer/berlinsparqlbenchmark/}} (BSBM)~\cite{bsbmshort}, Waterloo SPARQL Diversity Test Suite\footnote{\url{https://dsg.uwaterloo.ca/watdiv/}} (WatDiv)~\cite{watdivshort}, FishMark~\cite{fishmark}, BowlognaBench~\cite{blow}, and BioBench-Allie\footnote{\url{http://allie.dbcls.jp/}}~\cite{biobench}. WatDiv and BSBM follow specific rules that allow us to scale the datasets to arbitrary sizes using their scale factors while other datasets are of fixed siz. \reftab{table::kgs} shows the statistical information related to the above benchmarks. The RDF representations of these benchmarks are available in different formats such as N-Triples, Turtle, and XML. We used the RDF/N-Triples format. However, to load them into a document-store like MongoDB, we had to convert them to the JSON-LD\footnote{\url{https://www.w3.org/2018/jsonld-cg-reports/json-ld/}} format. We performed the conversion using a parser designed and developed as part of this project\footnote{The source code is available through \url{https://github.com/oursubmission/ESWC}}. %We modified some queries (e.g., by removing query modifiers such as union and order by clauses) wherever necessary because the main focus of our paper is on the different query types. Note that our approach to modifying queries is similar to that of other studies~\cite{Medha2,triad,weikumjoin}.

We ran the benchmark queries against the corresponding datasets using the four DMSs. We selected twelve queries across the benchmarks that were representative of the major four query types. These queries provide varying degrees of selectivity and complexity.\footnote{These queries are available through:\url{https://github.com/oursubmission/ESWC}} The selected \textbf{subject-subject join} queries are: Query 5 from FishMark (\textbf{FishMark-Q5}), Query 7 from BowlognaBench (\textbf{BowlognaBench-Q7}), and Query 7 from WatDiv (\textbf{WatDiv-Q7}). The selected \textbf{subject-object join} queries are: Query 2 from BioBench-Allie (\textbf{BioBench-Allie-Q2}), Query 21 from WatDiv (\textbf{WatDiv-Q21}), and Query 22 from WatDiv (\textbf{WatDiv-Q22}) and the selected \textbf{tree-like join} queries are: Query 1 from BioBench-Allie (\textbf{BioBench-Allie-Q1}), Query 14 from BowlognaBench (\textbf{BowlognaBench-Q14}), and Query 19 from FishMark (\textbf{FishMark-Q19}). Finally, the selected \textbf{optional join} queries are: Query 2 from BSBM (\textbf{BSBM-Q2}), Query 4 from BSBM (\textbf{BSBM-Q4}), and Query 2 from FishMark (\textbf{FishMark-Q2}). For WatDiv and BSBM benchmarks, corresponding query execution times over KGs with 100M triples are reported in this paper while results for 10M and 1000M only available online\footnote{These queries are available through:\url{https://github.com/oursubmission/ESWC}} due to space constraint.

%########################################
%       Settings
%########################################
\subsection{System Settings}
\label{sec:db}
\noindent \textbf{Computational Environment.} Our benchmark system is a Virtual Machine (VM) instance with a 2.3GHz AMD Processor, running Ubuntu Linux (kernel version: 4.4.0-170-generic), with 48GB of main memory, 16 vcores, 512K L2 cache, 5TB instance storage capacity. The VM cache read is roughly 2799.45MB/sec and the buffer read is roughly 35.85MB/sec (i.e., the output of the ``hdparm -Tt'' Linux command). The operating system is set with almost no ``soft/hard'' limit on the file size, CPU time, virtual memory, locked-in-memory size, open files, processes/threads, and memory size using Linux ``ulimit'' settings.

\noindent \textbf{Data Management Systems (DMSs).} We chose four different DMSs: (1) Row-store Virtuoso (Open Source Edition, version 06.01.3127), (2) Column-store Virtuoso (Open Source Edition, Version 07.20.3230--commit 4a668a5), (3) Blazegraph\footnote{Previously known as Bigdata DB.} (Open Source Edition, version 2.1.5--commit 3122706), and (4) MongoDB (community edition, version: 4.0.9).

\noindent \textbf{Configuration of row- and column-store Virtuoso.} We configured both of them based on the vendor's official recommendations.\footnote{\url{http://vos.openlinksw.com/owiki/wiki/VOS/VirtRDFPerformanceTuning}} For example, we configured the Virtuoso process to use the main memory and the storage disk effectively by setting ``NumberOfBuffers'' to ``4,000,000'', ``MaxDirtyBuffers'' to ``3,000,000'', and ``MaxCheckpointRemap'' to ``a quarter'' of the database size as recommended. We also used the latest version of GNU packages that are necessary to build column-store Virtuoso (e.g. GNU gpref 3.0.4, libtool 2.4.6, flex 2.6.0, Bison 3.0.4, and Awk 4.1.3).

\noindent \textbf{Configuration of Blazegraph.} We configured it based on the vendor's official recommendations\footnote{\url{https://wiki.blazegraph.com/wiki/index.php/PerformanceOptimization}} as well. For example, we turned off all inference, truth maintenance, statement identifiers, and the free text index in our experiment since reasoning efficiency was not part of our research focus in this paper.

\noindent \textbf{Configuration of MongoDB.} We used its default settings. We set its level of \textit{profiling to ``2''} to log the data for all query-related operations for precise and detailed query execution time extraction.

\noindent \textbf{Indexing of Virtuoso.} We did not change the default indexing scheme of \textit{Virtuoso} (both row- and column-store). As highlighted in the official website, ``alternate indexing schemes are possible but will not be generally needed''.\footnote{\url{http://docs.openlinksw.com/virtuoso/rdfperfrdfscheme}} More specifically, Virtuoso's data modeling is based on a relational table with three columns\footnote{In the case of loading named graphs, it adds another column for the context, called C.} for S, P, and O (i.e., S: Subject, P: Predicate, and O: Object) and carrying multiple indexes over that table to provide a number of different access paths. Most recently, column-store Virtuoso added columnar projections to minimize the on-disk footprint associated with RDF data storage. Virtuoso (both row- and column-store) creates the following compound indexes by default for the loaded KG: PSO, PO, SP, and OP.

\noindent \textbf{Indexing of Blazegraph.} As recommended in its official website,\footnote{\url{https://wiki.blazegraph.com/wiki/index.php/PerformanceOptimization}} we did not change its default indexing schema. In Blazegraph, indexes are based on \mbox{``B+Trees''} data structure. Blazegraph typically uses the following three indexes for triples modes: SPO, POS, and OSP. For normal use cases, these indexes are laid out on variable sized pages. These index pages are read from the backing store and load in the main memory on demand (i.e., into the Java heap). However, Blazegraph takes advantage of a variety of data structures to execute queries when stored KG content is loaded in the main memory. For example, the underlying data structure is retained by a mixture of a ring buffers on the stack alongside a native memory cache for buffering writes to reduce write application effects.

\noindent \textbf{MongoDB Storage Layouts.} We did not change its default storage engine which is a key/value store, namely, WiredTiger, to store JSON documents.\footnote{MongoDB uses the binary equivalent of each JSON document (i.e., BSON) for storage, in which the structure of each document remained unchanged.} MongoDB usually assigns an arbitrary (and unique) identifier to each JSON document as a key and considers the document as a value. It uses \mbox{B-Trees} to create indexes on the contents of each JSON document.

\noindent \textbf{Indexing of MongoDB.} We created indexes on those name/value pairs of the JSON-LD that were representatives of subjects and predicates.

\noindent \textbf{Loading the benchmark KG.} We loaded the RDF/N-Triples format of benchmarks into \textit{Virtuoso} (row- and column-store) by using its native bulk loader function (i.e., ``ld\_dir''). To load the KGs into \textit{Blazegraph}, we used its native ``DataLoader'' utility\footnote{\url{https://wiki.blazegraph.com/wiki/index.php/Bulk_Data_Load}}. We loaded KGs into MongoDB using its native tool, called ``mongoimport''.

\noindent \textbf{Shutdown store, clear caches, restart store.} We measured the query execution times in our evaluation. This is an end-to-end time computed from the time of query submission to the time when the result is outputted. After the execution of each query, we carefully checked to ensure that the output results are correct and exactly the same across different DMSs. The query times for both cold- and warm-run (aka, cold and warm cache) are reported. For cold-run we dropped the file systems caches using \texttt{/bin/sync}, \texttt{echo 3 > /proc/sys/vm/ drop\_caches}, and \texttt{swapoff -a \&\& swapon -a} commands. For fairness, the warm-run query times reported for each DMS are averaged\footnote{Geometric mean is used.} over 5 successive runs (with almost no delay in between).% to account for any randomness and noise.

\begin{figure}[t]
%\captionsetup[subfigure]{aboveskip=-2pt,belowskip=-2pt}
\centering
\begin{subfigure}{1\textwidth}
  \centering
  \includegraphics[width=0.8\linewidth]{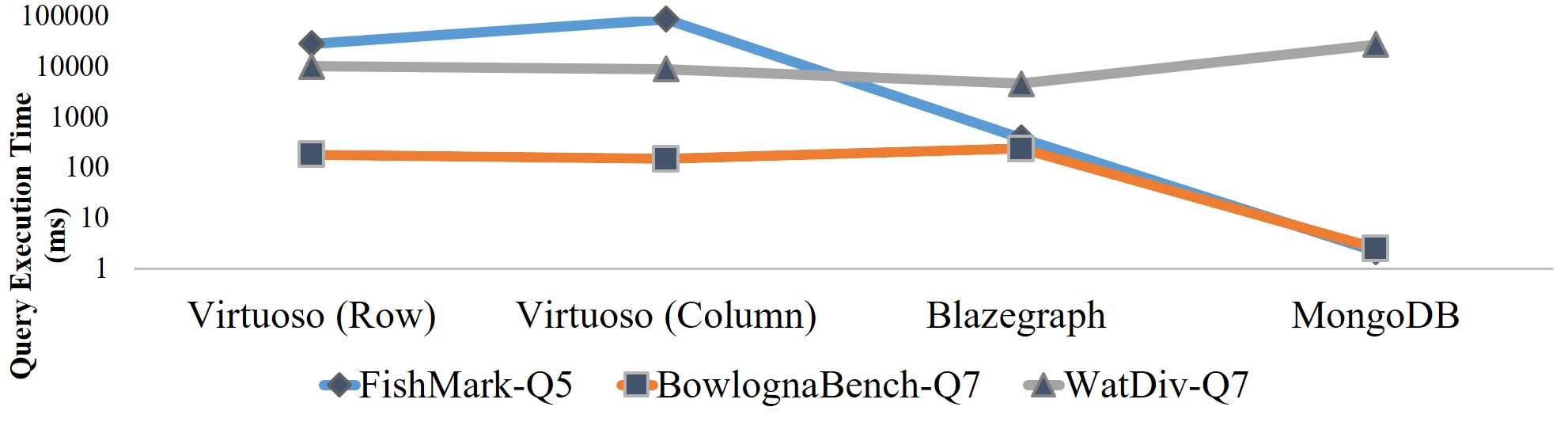}
  \caption{Subject-subject join (warm)} 
  %\label{fig:sub1}
\end{subfigure}%
\vspace*{.2cm}
\begin{subfigure}{1\textwidth}
  \centering
  \includegraphics[width=0.8\linewidth]{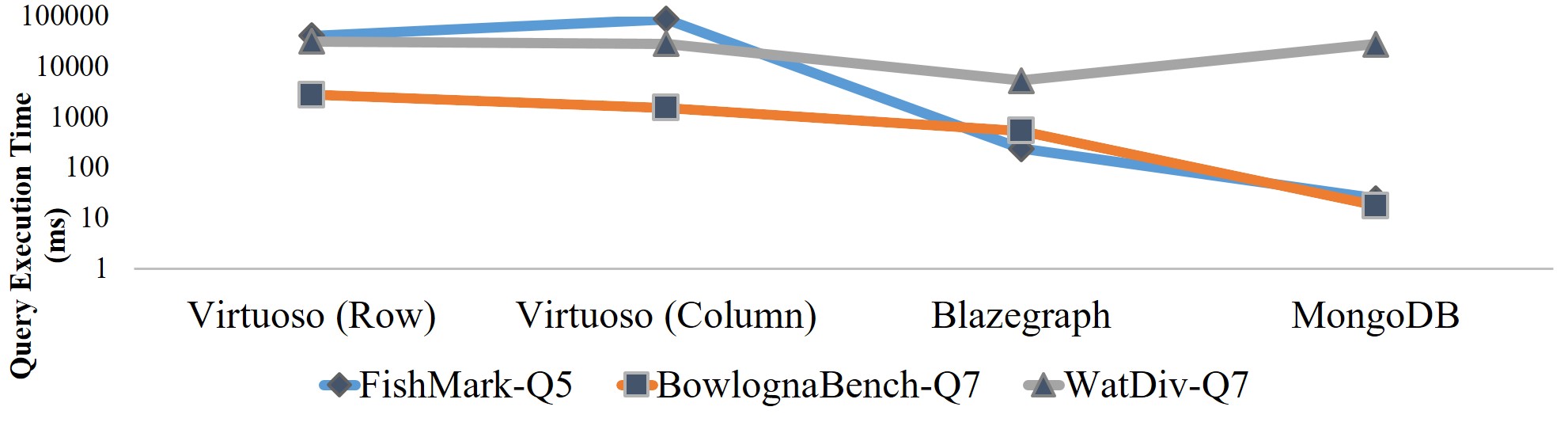}
  \caption{Subject-subject join (cold)}
  %\label{fig:sub2}
\end{subfigure}
\caption{Impacts of subject-subject join queries on the DMSs (cold- and warm-run). $X$ axis shows DMSs and $Y$ axis shows the execution time of each query in milliseconds (using log scale).}
\label{fig::ss}
\end{figure}

%###################################
%
%      Evaluation
%##################################
\section{Evaluation}
\label{sec::Evaluation}
%\noindent In this section, we present our experimental results and analyses separately for each query type.

\subsection{Results}
\label{sec::results}

\textbf{Subject-subject Joins.} The query execution times are shown in \reffig{fig::ss} in which $X$ axis represents the DMSs and the $Y$ axis shows the execution times of queries, namely, \textbf{FishMark-Q5}, \textbf{BowlognaBench-Q7}, and \textbf{WatDiv-Q7} in milliseconds (using log scale). \reffig{fig::ss} shows that MongoDB runs this type of queries over one order of magnitude faster than the other DMSs when queries are high-selective. For example, MongoDB executed \textbf{FishMArk-Q5} in 2.19 milliseconds (warm-run) while Blazegraph, column-store Virtuoso, and row-store Virtuoso executed the same query in 394.24, 89543.81, and 29045.86 milliseconds, respectively. However, our results show that Blazegraph performs at least 2x faster than other DMSs when subject-subject join queries are low-selective (e.g., \textbf{WatDiv-Q7}). In \reffig{fig::ss}, the differences between cold- and warm-run show that Virtuoso (row- and column-store) can take advantage of caching techniques more than other DMSs. For example, Virtuoso (row and column) executes \textbf{BowlognaBench-Q7} in over 1500 milliseconds (cold-run) while its execution time is around 150 milliseconds in a warm-run.

\begin{figure}[t]
\centering
\begin{subfigure}{1\textwidth}
  \centering
  \includegraphics[width=0.8\linewidth]{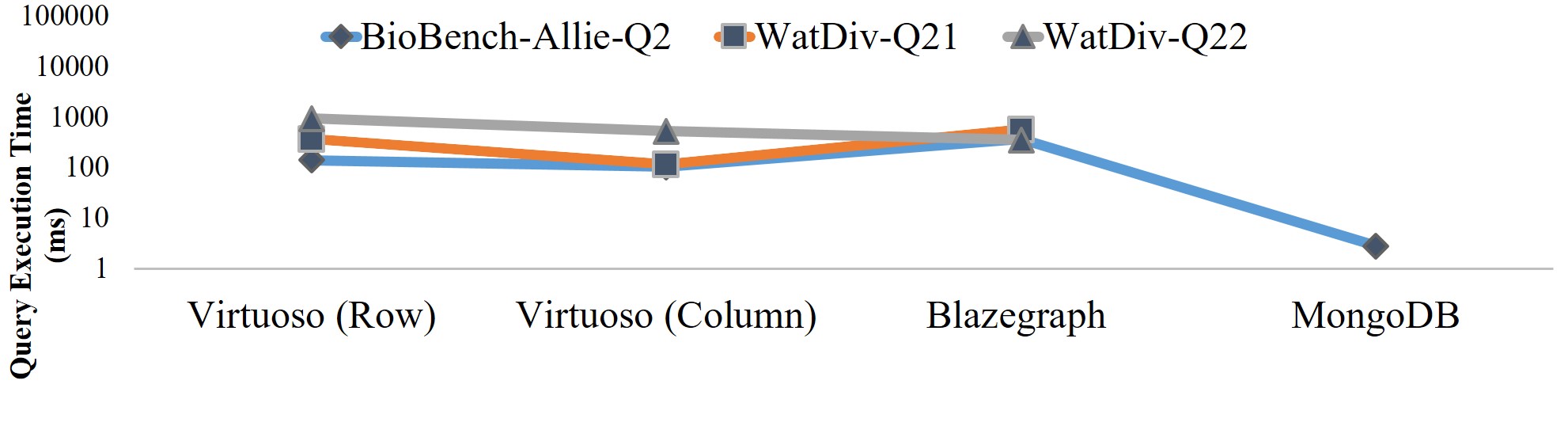}
  \caption{Subject-object join (warm)}
  %\label{fig:sub1}
\end{subfigure}%

\begin{subfigure}{1\textwidth}
  \centering
  \includegraphics[width=0.8\linewidth]{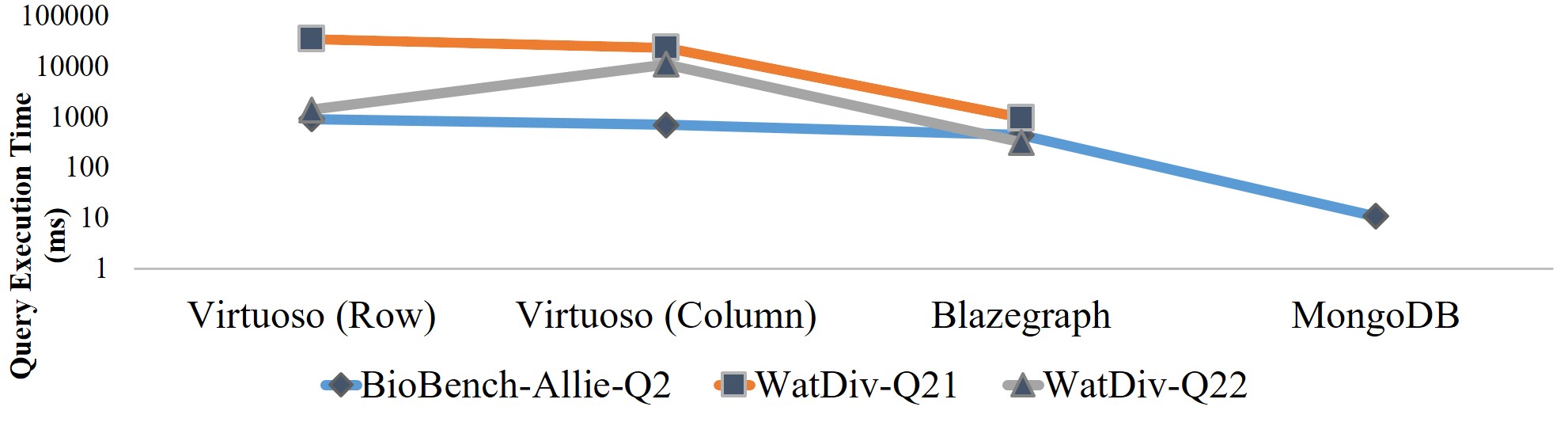}
  \caption{Subject-object join (cold)}
  %\label{fig:sub2}
\end{subfigure}
\caption{Impacts of subject-object join queries on the DMSs (cold- and warm-run).}
\label{fig::so}
\end{figure}

\textbf{Subject-object Joins.} The query execution times for the selected subject-object join queries is shown in \reffig{fig::so} in which $X$ axis shows the DMSs and the $Y$ axis shows the execution time of queries, namely, \textbf{BioBench-Allie-Q2}, \textbf{WatDiv-Q21}, and \textbf{WatDiv-Q22} in milliseconds (using log scale). Although MongoDB executed \textbf{BioBench-Aliie2-Q2}, which is a high-selective query, over 2 orders of magnitude faster than other DMSs, it could not finish the execution of \textbf{WatDiv-Q21} and \textbf{WatDiv-Q22} within the given time-out period of 50,000 milliseconds. The complexity and non-selectivity of these two queries may have contributed to the unsuccessful execution over MongoDB. However, \reffig{fig::so} shows that other DMSs performed comparably. For instance, \textbf{WatDiv-Q21} executed over Blazegraph in around 570 milliseconds (warm-run) where this execution time is equal to 118.38 and 374.6 milliseconds for column-store Virtuoso and row-store Virtuoso, respectively.

%We speculate that Blazegraph's implementation of $B^+$-tree default indexing schema (inspired by Google BigTable) and running of operators on Java Virtual Machine (JVM) heap contributed to the faster query execution time for \textbf{BSBM-Q7}.

\begin{figure}[t]
\centering
\begin{subfigure}{1\textwidth}
  \centering
  \includegraphics[width=0.8\linewidth]{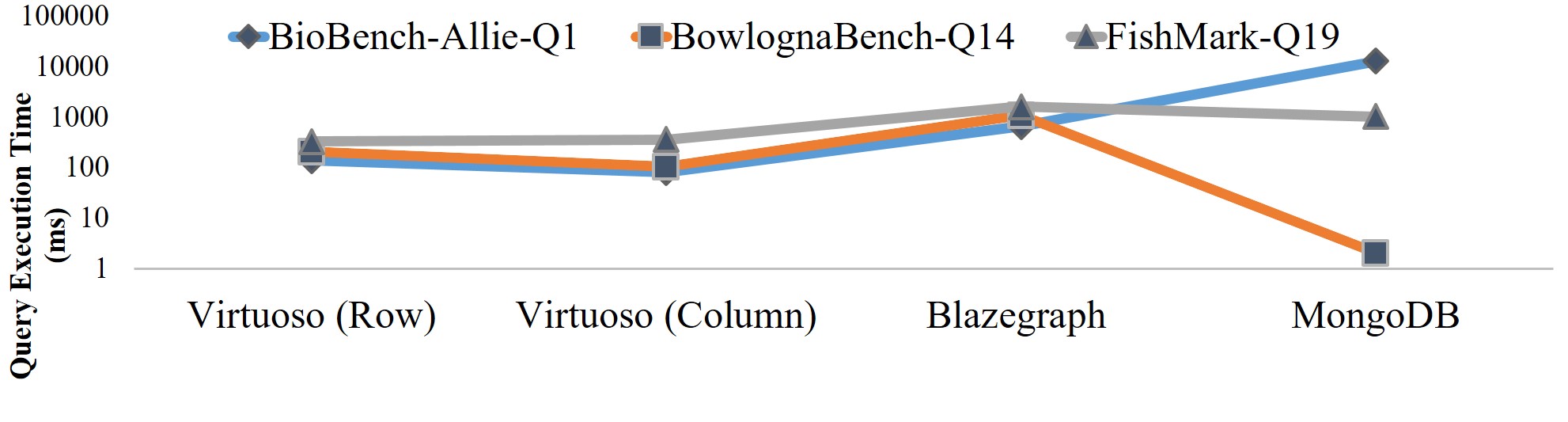}
  \caption{Tree-like join (warm)}
  %\label{fig:sub1}
\end{subfigure}%

\begin{subfigure}{1\textwidth}
  \centering
  \includegraphics[width=0.8\linewidth]{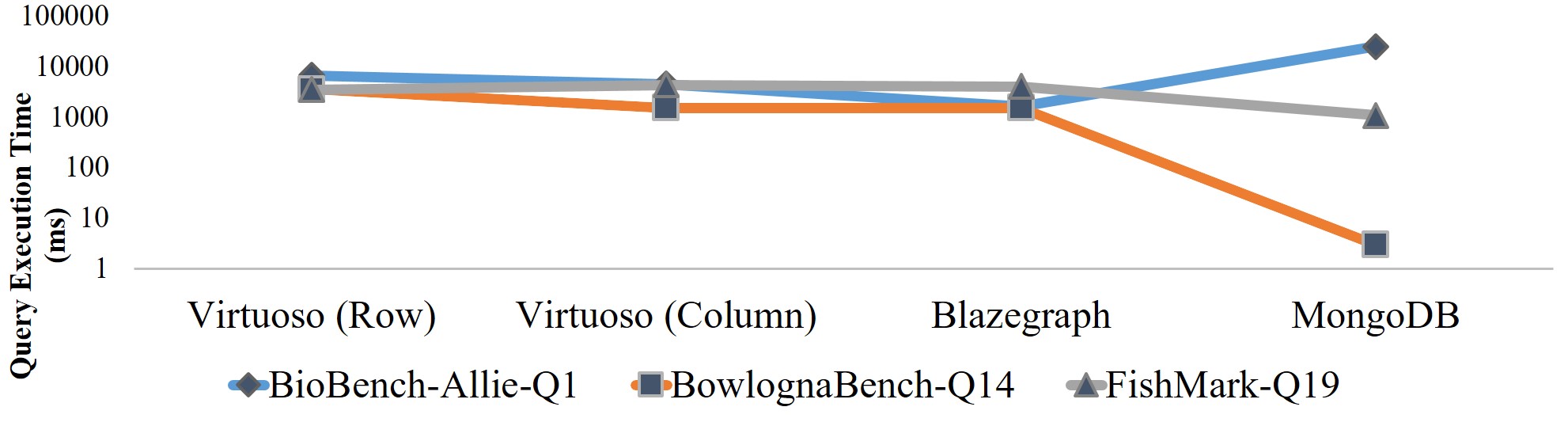}
  \caption{Tree-like join (cold)}
  %\label{fig:sub2}
\end{subfigure}
\caption{Impacts of tree-like join queries on the DMSs (cold- and warm-run).}
\label{fig::tree}
\end{figure}

\textbf{Tree-like Joins} The query execution time for the selected tree-like join queries is shown in \reffig{fig::tree}. This figure shows that row- and column-store Virtuoso performed similarly for warm-run execution of \textbf{BioBench-Allie-Q1} and \textbf{FishMark-Q19} while Blazegraph is around 5x slower. MongoDB appeared to be the slowest for warm-run execution of \textbf{BioBench-Allie-Q1} while its performance is comparable with Blazegraph for \textbf{FishMark-Q19}. \textbf{BowlognaBench-Q14} executed around 2 orders of magnitude faster using MongoDB, probably because it is a high-selective tree-like join query. The comparison between cold- and warm-run execution of \textbf{FishMark-Q19} can also give rise to the importance of the role that caching techniques play in query performance where MongoDB is the fastest in cold-run, but in warm-run, it is almost the slowest (after Blazegraph).

\begin{figure}[t]
\centering
\begin{subfigure}{1\textwidth}
  \centering
  \includegraphics[width=0.8\linewidth]{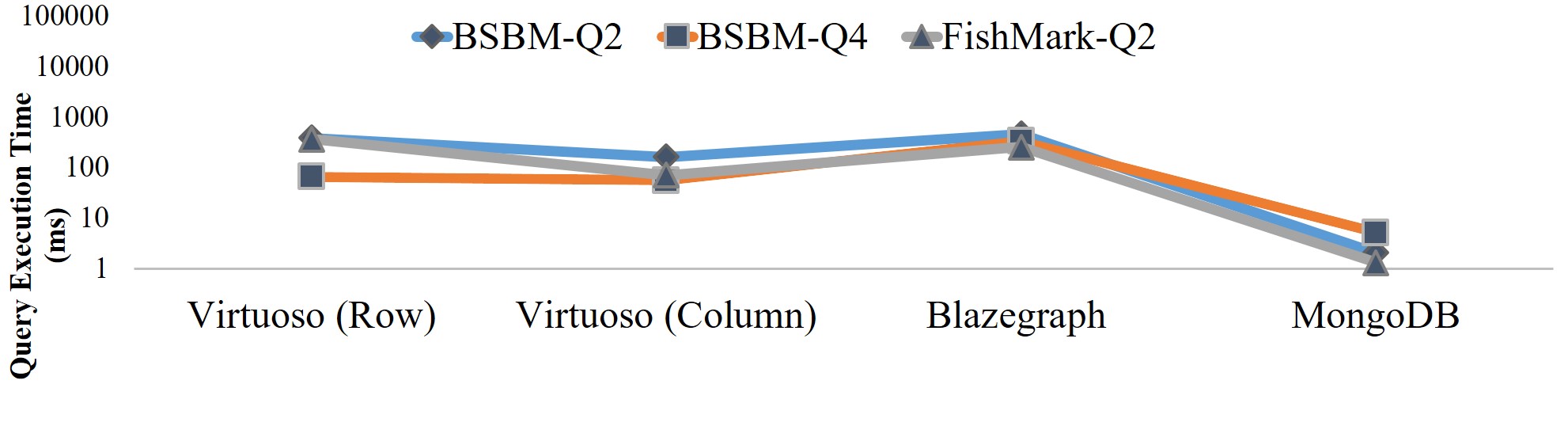}
  \caption{Optional join (warm)}
  %\label{fig:sub1}
\end{subfigure}%

\begin{subfigure}{1\textwidth}
  \centering
  \includegraphics[width=0.8\linewidth]{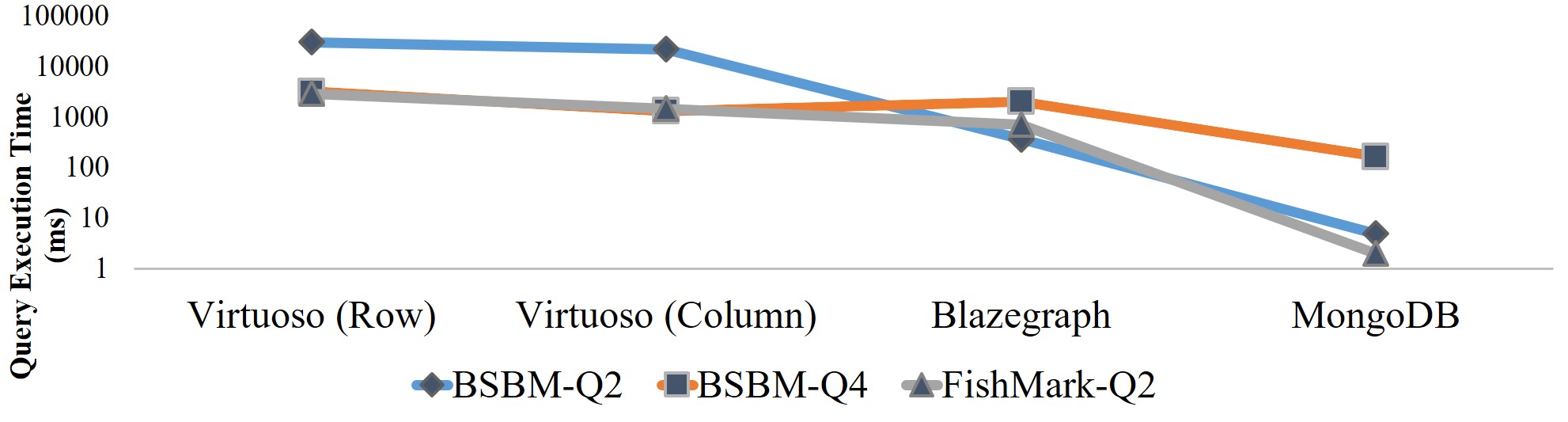}
  \caption{Optional join (cold)}
  %\label{fig:sub2}
\end{subfigure}
\caption{Impacts of optional join queries on the DMSs (cold- and warm-run).}
\label{fig::opt}
\end{figure}

\textbf{Optional Joins.} The query execution time for the selected optional join queries is shown in \reffig{fig::opt} in which MongoDB executed them faster than other DMSs. High-selectivity of the selected queries may have been an important factor in MongoDB's performance advantage. Row-store Virtuoso was the slowest across others while column-store Virtuoso performed over 3x faster than Blazegraph to run these queries (warm-run). However, in the cold-run, aside from the performance advantage of MongoDB, Blazegraph performed slightly better than others especially for executing \textbf{BSBM-Q2}.

\subsection{Analysis}
\label{sec::analysis}
%In order to explain the key factors contributing to the observed performance differences, we present our detailed analyses below.

\textbf{Subject-subject Join.} Our results showed that MongoDB can execute this query type over one order of magnitude faster, especially for queries with high selectivity. Virtuoso (both row- and column-store) and Blazegraph typically execute subject-subject join by scanning indexes for each triple pattern separately. The retrieved result of each triple pattern is kept in the main memory as an intermediary result. These DMSs join different intermediary results to return the final result. Virtuoso (both row- and column-store) and Blazegraph typically use a hash join algorithm for executing subject-subject joins over the intermediary results. However, in MongoDB, all triples with the same subject have appeared in a single JSON document and the joining of triples with the same subject is equivalent to an index-based look-up querying of a given subject. Therefore, we observed better performance from MongoDB for high-selective subject-subject join queries.

\textbf{Subject-object Joins.} We observed that Blazegraph offers a significant performance improvement on the cold-run execution of subject-object join queries. Merge join is known to be an efficient algorithm to be implemented by DMSs for running subject-object join over intermediary results, after scanning indexes, to return the final result~\cite{svenbook}.\footnote{Please note that~\cite{svenbook} did not use the exact term ``subject-object join'', instead it refers to this query type by its definition} To the best of our knowledge, none of the DMSs have implemented the merge join as a part of their query processing engines. As a result, these DMSs use Index Nested Loop join algorithm to support subject-object queries. In our experiments, the faster query execution time of Blazegraph (cold-run) for this query type may stem from the use of $B^+$-tree-based index nested loop join which is more read-optimal as compared to bitmap index-based of both row- and column-store Virtuoso. In addition, Blazegraph uses cardinality estimation to predict the size of the intermediary results of queries to find a good join ordering. This estimation requires dynamic programming techniques and the building of statistical summaries such as histograms. Such cardinality estimation has a significant performance effect on the execution time of low-selective subject-object join queries. Therefore, we observed better performance from Blazegraph as compared to row- and column-store Virtuoso for cold-run execution of subject-object join queries.

\textbf{Tree-like Joins.} Our results showed that Virtuoso (specially column-store) executed tree-like join queries faster. Technically, a tree-like join can be considered as a combination of subject-subject and subject-object joins. The performance of tree-like query types may vary depending on the complexity of such a combination and the efficiency of DMSs' query optimizer. In our experiments, Virtuoso (row- and column-store) showed better performance for executing this query types with lower selectivity. We speculate that Virtuoso's vectorized query execution model and its secondary indexing strategies (aka, compound indexes) along with its well-engineered query optimization engine may explain the better performance as compared to others. In addition, column-store Virtuoso usually stores indexes more compactly. Therefore, it can store and index short, fixed-length identifiers rather than string literals of subject, predicate, and object values. This compactness typically contributes to faster index selection in query planning and has a positive performance impact on tree-like join queries.

\textbf{Optional Joins.} There is no evidence that any of the four DMSs has implemented specialized optimizations for optional joins. As a result, although MongoDB showed better performance for running high-selective queries, we do not observe significant differences in query performance among the DMSs for low-selective queries. We note that both row- and column-store Virtuoso come with a compression strategy for storing KG datasets. Furthermore, bitmap indexing provides row- and column-store Virtuoso with better space utilization as compared to $B^+$-tree of Blazegraph (or MongoDB). In this regard, we speculate Virtuoso is likely to aggressively prune intermediate results and perform faster than others for optional join query processing, especially for low-selective optional join queries.

\textbf{Scale effects.} FishMark, BioBench-Allie, and BowlognaBench are fixed-size datasets that cannot be scaled. In contrast, WatDiv and BSBM are scalable. In this paper, we reported the corresponding query execution times of these two datasets with 100M triples. However, corresponding results for datasets with 10M and 1000M are computed and are available online.\footnote{\url{https://github.com/oursubmission/ESWC}} Our results showed that in most cases, selectivity and query type along with query optimization and caching techniques are probably more significant contributory factors to the performance differences across employed DMSs as compared to the size of the underlying dataset (i.e., the scale factor).

%The foregoing explanation suggests that no single DMS displays superior query performance across different query types.

%As we speculated in \refsec{sec::queryexpectation}, Virtuoso performs around one order of magnitude faster than others for executing this type of queries, especially at a lower scale. However, after executing \textbf{BSBM-Q4}, we would amend our previous speculation slightly by highlighting that in addition to bitmap indexing and data compression, the cardinality of resultset has an impact on the optional join queries. A summary of our findings is presented in \reftab{table::summary}.

%------------------------------------
%           Discussion
%------------------------------------
\section{Discussion}
\label{sec::Discussion}

\subsection{Limitations}
Our results indicate that no single DMS displays superior query performance across different query types. These results are likely to be generalizable. However, more experimentation is warranted before we can arrive at any firm conclusions. In our experiments, we had four archetypal query types. However, there may be other query types that we need to consider in the future.

Currently, the maximum size of each JSON document in MongoDB is 16MB. It rejects JSON documents when its size exceeds this value. Technically, the maximum document size in document-stores helps ensure that a single document cannot use an excessive amount of memory, but the JSON-LD representation of KGs might be affected negatively by this. In our experiments, there were no cases in which the document size exceeded the maximum value. However, in principle, the size of JSON documents may exceed the maximum document size depending on the KG content.

Another issue that remains to be addressed is the automatic conversion of SPARQL to JavaScript-like (i.e., for MongoDB) queries. In our experiments, we converted the benchmark queries manually and after the execution of each query, we carefully checked to ensure that the output results are correct and exactly the same across different DMSs and representations.

\subsection{Lessons Learned}

%\textbf{Lack of Schema and Heterogeneity of the Underlying Data.} KGs can represent and mix diverse data ranging from \textit{structured} (e.g., BSBM) to \textit{unstructured} (e.g., WatDiv). This \textit{structural flexibility} can contribute to the widespread acceptance of KGs in a variety of domains in real-world. However, this structural mixture and flexibility may pose a challenge to DMSs for \textit{querying} KG content efficiently since DMSs typically cannot make any \mbox{priori} assumptions about the structure of the data that is going to be stored. The lack of schema and heterogeneity of the underlying data makes the execution of different KG query types a hard problem.

\textbf{Intermediary Result.} We note that the performance of different query types tends to be negatively affected by the sizes of the query's output and more often its intermediary results. When a query type contains more than a triple pattern, DMSs usually have to scan large parts of indexes for each triple pattern and then join the result of these scans. These index scans would produce large intermediary results. We observed that even when the query itself is very selective with small output, the size of the intermediary results can be still very large. The size of the intermediary result challenges DMSs. Currently, DMSs usually use either of two techniques data compression or Sideways Information Passing (SIP) to decrease the size of intermediary results. It appears that employing these techniques to decrease intermediary results may increase the computation need of the query evaluation process for the uncompression or additional filtering (for SIP) requirement.

\textbf{Locality.} Column-store Virtuoso and MongoDB are designed to increase data locality while storing KGs' content more than others. In the column-store Virtuoso storage model, each column of a table or index is stored contiguously to provide physical adjacency. Therefore, when queries (e.g., tree-like joins) need to access a subset of columns from one table, only those columns actually being accessed need to be read from disk which can be culminated to better use of I/O throughput and memory. This locality has this potential to reduce the traffic between CPU cache and main memory and provide a better CPU utilization. MongoDB similarly takes advantage of data locality since all the triples related to one resource (i.e., a subject in the JSON-LD) are physically located together. We speculate that such locality leads to denser data layout, more CPU cache (i.e., L2 cache) locality and more RAM locality and therefore increased overall performance on high-selective KG queries.

\textbf{Cache Efficiency.} DMSs usually utilize their internal and the underlying filesystem cache memory. When enough free memory available and allocated to DMSs, efficient utilization of this memory for caching purposes can typically contribute to faster warm-run query execution. Comparing the results of different queries across the DMSs in cold- and warm-run query execution suggested that column-store Virtuoso provides better cache management. In applications with ad-hoc queries, the cache management may not impact the performance significantly, but for cases in which a number of queries are repeated periodically, employing suitable cache techniques can positively contribute to query performance.

\section{Related Work}
\label{sec::related-work}
%\noindent Efficient data management plays an important role in unlocking the full potential of Semantic Web applications. DMS designers have employed a variety of \textit{design choices and architectures} to present efficient systems for these applications. 

Early approaches employed relational database systems to store the Semantic Web datasets. In addition, several approaches have exploited NoSQL databases for building DMSs as discussed in~\cite{Survey2018short}. We can classify these studies into three categories: \textbf{Triple-based Indexing.} HexaStore~\cite{Hexastore} is a well-known DMS based on indexing. This created indexes on all permutations of the triple pattern. The effectiveness of triple-based indexing solutions can be limited since querying KGs typically requires touching a large amount of data and complex filtering. \textbf{Infrastructure Configuring.} JenaHBase~\cite{JenaHbase} and H2RDF~\cite{H2rdf} are well-known DMSs that focused mainly on the configurations of underlying infrastructure such as cluster segmentation, communication overhead, and distributed storage layouts. \textbf{Graph Processing.} Graph-based stores usually model Semantic Web data as a labeled and directed multi-edge graph by using a disk-based adjacency list table and executes queries by mapping them to a sub-graph matching task over the graph. %Nonetheless, previous studies generally focused on the importance of cluster size, communication efficiency, database tuning, etc. for improving the performance of query execution for Semantic Web applications.

In addition to the design of the DMSs, analysis of available DMSs using benchmark datasets has been a core topic of Semantic Web data management research. For example, some studies such as~\cite{bsbmshort,watdivshort,ldbcshort} presented new benchmark datasets. Some other studies such as~\cite{ISWC2013short} did not propose any new dataset, but tried to use available benchmarks and DMSs for the same purpose such as reporting key advantages and drawbacks of each DMS. There are also studies such as~\cite{Saleem} which comprehensively surveyed and analyzed available datasets in terms of different metrics such as the number of projection variables, the number of BGPs, etc. However, to the best of our knowledge, our paper is one of the first that investigated the comparative KG query performance by mapping archetypal SPARQL query types with different DMS types.

\section{Conclusion}
\label{sec::conclusion}
\noindent We have focused on the mapping of different types of KG queries onto different types of DMS. We analyzed the performance of row-store Virtuoso, column-store Virtuoso, Blazegraph (i.e., graph-store), and MongoDB (i.e., document-store) using five well-known benchmarks, namely, BSBM, WatDiv, FishMark, BowlognaBench, and BioBench-Allie. A summary of our findings is as follows:

\begin{itemize}
    \item There are significant interaction effects between different types of DMSs and query types.
    
    \item Our results showed that the simplicity of the underlying storage layout, increasing data locality, and suitable caching techniques in \textit{Virtuoso} (specially column-store) lead to a performance advantage for tree-like join queries by generating \textit{smaller intermediary results}.
    
    \item We also found that suitable \textit{cardinality estimation} as well as efficient query optimization of \textit{Blazegraph} offers a significant performance improvement on the cold-run execution of subject-object join queries.
    
    \item Taking advantage of data locality and employing efficient data structures such as B-trees for implementing indexes in MongoDB can contribute to over one order of magnitude better performance for executing subject-subject join queries, especially for queries with high selectivity.

\end{itemize}

The results presented in this paper can assist in the benchmarking of the emerging type of DMSs. However, more experimentation is warranted before we can arrive at any firm conclusions. In addition, our experience while performing a comparative analysis of KG query performance raised several new and interesting questions and research directions that need to be addressed in the future. These include replication of this research using more datasets and DMSs and automatic rewriting of SPARQL queries to other declarative query languages such as MongoDB's query language.

%-----------------------------------------------
% References
%-----------------------------------------------

\bibliographystyle{abbrv}
\bibliography{main}

\begin{thebibliography}{10}

\bibitem{tamer2019}
G.~Alu\c{c}, M.~T. \"{O}zsu, and K.~Daudjee.
\newblock Building self-clustering rdf databases using tunable-lsh.
\newblock {\em The VLDB Journal}, 28(2):173--195, 2019.

\bibitem{watdivshort}
G.~{Alu{\c{c}} et. al.}
\newblock Diversified stress testing of {RDF} data management systems.
\newblock In {\em Proc. of the Int. {Semantic Web} Conf. (ISWC)}, pages
  197--212, 2014.

\bibitem{Medha2}
M.~Atre.
\newblock Left bit right: For {SPARQL} join queries with {OPTIONAL} patterns
  (left-outer-joins).
\newblock In {\em Proc. of the ACM Int. Conf. on Management of Data (SIGMOD)},
  pages 1793--1808, 2015.

\bibitem{fishmark}
S.~{Bail et. al.}
\newblock {FishMark}: A linked data application benchmark.
\newblock In {\em Proc. of the Int. Workshop on Scalable and High-Performance
  Semantic Web Systems (SSWS)}, pages 1--15, 2012.

\bibitem{bsbmshort}
C.~{Bizer et. al.}
\newblock The {Berlin SPARQL} benchmark.
\newblock {\em Int. J. {Semantic Web} Inf. Syst.}, 5:1--24, 2009.

\bibitem{ISWC2013short}
P.~{Cudr{\'e}-Mauroux et. al.}
\newblock {NoSQL} databases for {RDF}: An empirical evaluation.
\newblock In {\em Proc. of the Int. {Semantic Web} Conf. (ISWC)}, pages
  310--325, 2013.

\bibitem{blow}
G.~{Demartini et. al.}
\newblock Bowlognabench---benchmarking rdf analytics.
\newblock In {\em Proc. of of the Int. Symp. on Data-Driven Process Discovery
  and Analysis (SIMPDA)}, pages 82--102, 2012.

\bibitem{IBMapple}
S.~{Duan et. al.}
\newblock Apples and oranges: A comparison of {RDF} benchmarks and real {RDF}
  datasets.
\newblock In {\em Proc. of the ACM Int. Conf. on Management of Data (SIGMOD)},
  pages 145--156, 2011.

\bibitem{ldbcshort}
O.~{Erling et. al.}
\newblock The {LDBC} social network benchmark: Interactive workload.
\newblock In {\em Proc. of the ACM Int. Conf. on Management of Data (SIGMOD)},
  pages 619--630, 2015.

\bibitem{svenbook}
S.~Groppe.
\newblock {\em Data management and query processing in {Semantic Web}
  databases}.
\newblock Springer Science \& Business Media, 2011.

\bibitem{JenaHbase}
V.~{Khadilkar et. al.}
\newblock Jena-hbase: A distributed, scalable and efficient {RDF} triple store.
\newblock In {\em Proc. of the Int. {Semantic Web} Conf. (ISWC)}, pages 85--88,
  2012.

\bibitem{RDF3x}
T.~Neumann and G.~Weikum.
\newblock The {RDF-3X} engine for scalable management of {RDF} data.
\newblock {\em Proc. VLDB Endow.}, 19(1):91--113, 2010.

\bibitem{H2rdf}
N.~Papailiou, I.~Konstantinou, D.~Tsoumakos, and N.~Koziris.
\newblock {H2RDF}: Adaptive query processing on {RDF} data in the cloud.
\newblock In {\em Proc. of the Int. Conf. on World Wide Web (WWW)}, pages
  397--400, 2012.

\bibitem{kgsurvey}
H.~Paulheim.
\newblock Knowledge graph refinement: {A} survey of approaches and evaluation
  methods.
\newblock {\em Semantic Web}, 8(3):489--508, 2017.

\bibitem{Saleem}
M.~{Saleem et. al.}
\newblock How representative is a {SPARQL} benchmark? an analysis of {RDF}
  triplestore benchmarks.
\newblock In {\em Proc. of the Int. Conf. on World Wide Web (WWW)}, pages
  1623--1633, 2019.

\bibitem{selectivity}
M.~{Stocker et. al.}
\newblock Sparql basic graph pattern optimization using selectivity estimation.
\newblock In {\em Proc. of the Int. Conf. on World Wide Web (WWW)}, pages
  595--604, 2008.

\bibitem{Hexastore}
C.~{Weiss et. al.}
\newblock Hexastore: Sextuple indexing for {Semantic Web} data management.
\newblock {\em Proc. VLDB Endow.}, 1(1):1008--1019, 2008.

\bibitem{biobench}
H.~{Wu et. al.}
\newblock {BioBenchmark Toyama} 2012: an evaluation of the performance of
  triple stores on biological data.
\newblock {\em Journal of Biomedical Semantics}, 5(1):32--43, 2014.

\bibitem{Survey2018short}
M.~{Wylot et. al.}
\newblock {RDF} data storage and query processing schemes: A survey.
\newblock {\em ACM Comput. Surv.}, 51(4):84:1--84:36, 2018.

\end{thebibliography}

\end{document}